\documentclass[prb,preprintnumbers,amsmath,amssymb,floatfix]{revtex4}

\usepackage{graphicx}
\usepackage{subfigure}
\usepackage{dcolumn}

\begin{document}

\title{The linear and nonlinear Jaynes-Cummings model for the multiphoton transition}
\author{Xiao-Jing Liu$^{a,b}$, Jing-Bin Lu$^{a}$\footnote{E-mail:
lujbjlu@163.com}, Si-Qi Zhang$^{a}$, Ji-Ping Liu$^{b}$, Hong
Li$^{c}$, Yu Liang$^{b}$, Ji Ma$^{b}$ and Xiang-Yao Wu$^{b}$}
\affiliation{$^{a}$ \small Institute of Physics, Jilin University,
Changchun
130012 China \\{$^{b}$\small Institute of Physics, Jilin Normal University, Siping 136000, China}\\
$^{c}$ {\small Institute of Physics, Northeast Normal University,
Changchun 130012 China}}

\begin{abstract}
With the Jaynes-Cummings model, we have studied the atom and light
field quantum entanglement of multiphoton transition, and
researched the effect of initial state superposition coefficient
$C_{1}$, the transition photon number $N$, the quantum discord
$\delta$ and the nonlinear coefficient $\chi$ on the quantum
entanglement degrees. We have given the quantum entanglement
degrees curves with time evolution, and obtained some results,
which should have been used in quantum computing and quantum
information.

\vskip 10pt


\end{abstract}

\vskip 10pt \maketitle {\bf 1. Introduction} \vskip 10pt

The interaction between a two-level system and a harmonic
oscillator is ubiquitous in different physical setups, ranging
from quantum optics to condensed matter and applications to
quantum information. Typically, due to the parameter accessibility
of most experiments, the rotating-wave approximation (RWA) can be
applied producing a solvable dynamics called the Jaynes-Cummings
model (JCM) [1], this model has been widely generalized to treat
various interactions between atoms and photons. These include,
e.g., multilevel atoms interact with multimode quantized fields,
and various multiphoton processes in quantum optics [2]. It gives
arise to many quantum phenomena that can not be explained in
classical terms, such as the collapses and revivals of the atomic
population inversion [3], squeezing of the field [4], and
atom-cavity entanglement [5]. Recent experiments with Rydberg
atoms and microwave photons in a superconducting cavity have
turned the JCM from a theoretical curiosity to a useful and
testable enterprise [6]. Such a system is also suitable for
quantum state engineering and quantum information processing.

Entanglement plays a central role in quantum information, quantum
computation and communication, and quantum cryptography. A lot of
schemes are proposed for many-particle entanglement generation.
The simplest scheme to investigate the atom-field entanglement is
the Jaynes-Cummings model (JCM) describing an interaction of a
two-level atom with a singlemode quantized radiation field. This
model is of fundamental importance for quantum optics [7, 8] and
is realizable to a very good approximation in experiments with
Rydberg atoms in high-Q superconducting cavities, trapped ions,
superconducting circuits etc. [9, 10]. The model predicts a
variety of interesting phenomena. The atom-field entanglement is
among them. An investigation of the atom-field entanglement for
JCM has been initiated by Phoenix and Knight [11] and
Gea-Banacloche [12]. Gea-Banacloche has derived an asymptotic
result for the JCM state vector which is valid when the field is
initially in a coherent state with a large mean photon number.

The atomic systems have found new application in quantum
information processing [13]. The main requirement in quantum
information processing task is the quantum entanglement. The
interaction of atoms with cavity field has been shown to be an
efficient source of atom-atom, atom-cavity, and cavity-cavity
entanglement [14-16]. Appreciable number of studies related to the
study of influence of atom-field interaction on the entanglement
between atom and field has been done. The reason behind the study
of quantum entanglement in such system is due to its fundamental
nature and its applicability. However as discussed in previous
chapter quantum entanglement do not exhausts the complete quantum
correlations

The theory outlined in has been generalized for two-photon JCM
[17], two-photon JCM with nondegenerate two-photon and Raman
transitions [18, 19], two-atom JCM [20], two-atom one-mode Raman
coupled model [21] and two-atom two-photon JCM [22, 23].
Two-photon processes are known to play a very important role in
atomic systems due to high degree of correlation between emitted
photons. An interest for investigation of the two-photon JCM is
stimulated by the experimental realization of a two-photon
one-atom micromaser on Rydberg transitions in a microwave cavity
[24]. JCM with nondegenerate two-photon transitions have attracted
a great deal of attention. The foregoing model have been
considered in terms of atomic population dynamics research, field
statistics research, field and atom squeezing analysis, atom and
field entropy and entanglement examining [25]. The two-atom
two-photon JCM for initial two-mode coherent cavity field has been
investigated for nondegenerate two-photon transitions in [26].

The single photon and double photon Jaynes-Cummings model had been
studied largely. In this paper, we have studied the atom and light
field quantum entanglement of multiphoton transition, and
researched the effect of initial state superposition coefficient
$C_{1}$, the initial photon number $n$, the transition photon
number $N$ ($N=1, 2, 3, 4, 5, 6$), the quantum discord $\delta$
and nonlinear coupling constant $\chi$ on the quantum entanglement
degrees. We have given the quantum entanglement degrees curves
with time evolution, and obtained some results. When the
transition photon number $N$ increases, the entanglement degrees
oscillation get faster. When the initial state superposition
coefficient $C_{1}=0$,  with the quantum discord $\delta$
increase, the entanglement degrees oscillation get slowly. When
the initial state superposition coefficient $C_{1}=0.76$, the the
quantum discord increases, the entanglement degrees oscillation
get quickly. when the nonlinear coefficient $\alpha >0$, the
entanglement degrees oscillation get quickly. When the nonlinear
coefficient $\alpha <0$, the entanglement degrees oscillation get
slow. It is benefit to the atom and light field entanglement
obviously. These results have important significance in the
quantum communication and quantum information.

\vskip 10pt \maketitle {\bf 2. The multiphoton Jaynes-Cummings
model and entanglement degrees} \vskip 10pt

Let us consider the N-photon Jaynes-Cummings model, the
Hamiltonian is [27, 28]

\begin{eqnarray}
H=\omega
a^{+}a+\frac{1}{2}\omega_{0}\sigma_{z}+g({a^{+}}^{N}\sigma_{-}+a^{N}\sigma_{+})+\chi{a^{+}}^{2}a^{2},\hspace{0.2in}(\hbar=1)
\end{eqnarray}

where $a(a^{+})$ is the annihilation (creation) operator for a
photon in an electromagnetic mode of frequency, $¦Ø$ is the
radiative field mode frequency, $¦Ø_(0)$ is the atomic frequency.
, $g$ is field-atom coupling constant and $\sigma_{z}=\mid
a><a\mid-\mid b><b\mid$, $\sigma_{+}=\mid a><b\mid$,
$\sigma_{-}=\mid b><a\mid$. The nonlinear part
$\chi{a^{+}}^{2}a^{2}$ may be thought of as having been obtained
from a quartic potential through the rotating-wave approximation,
$\chi$ denotes the third-order susceptibility of Kerr medium, it
is the nonlinear coupling constant of light field and nonlinear
medium. This Hamiltonian, simple as it may seem, appears to lie in
the center of many theoretical investigations. Drummond and Walls
[29] and then Risken et al. [30] used it to examine the optical
bistability in nonlinear media [31,32].

The initial state is
\begin{eqnarray}
\mid\psi(0)>=c_{1}(0)\mid a,n>+c_{2}(0)\mid b,n+N>
\end{eqnarray}
where $|c_{1}(0)|^{2}+|c_{2}(0)|^{2}=1$, the state $\mid b>$ is
atom ground state, state $\mid a>$ is atom excited state, and the
wave function at any time is
\begin{eqnarray}
\mid\psi(t)>=c_{1}(t)\mid a,n>+c_{2}(t)\mid b,n+N>,
\end{eqnarray}
substituting Eqs. (1) and (3)into Schrodinger equation
\begin{eqnarray}
i\frac{\partial}{\partial t}\mid\psi(t)>=H\mid\psi(t)>,
\end{eqnarray}
we obtain
\begin{eqnarray}
&&i\frac{\partial}{\partial t}(c_{1}(t)\mid a,n>+c_{2}(t)\mid
b,n+N>)\nonumber\\&&=(\omega
a^{+}a+\frac{1}{2}\omega_{0}\sigma_{z}+g(a^{+N}\sigma_{-}+a^{N}\sigma_{+})+\chi{a^{+}}^{2}a^{2})(c_{1}(t)\mid
a,n>+c_{2}(t)\mid b,n+N>)\nonumber\\&&=(\omega
nc_{1}(t)+\frac{1}{2}\omega_0c_{1}(t)+gc_{2}(t)\sqrt{(n+1)(n+2)\cdot\cdot\cdot(n+N)}+\chi
n(n-1)c_{1}(t))\mid n,a>\nonumber\\&&+(\omega
(n+N)c_{2}(t)-\frac{1}{2}\omega_0c_{2}(t)+gc_{1}(t)\sqrt{(n+1)(n+2)\cdot\cdot\cdot(n+N)}\nonumber\\&&+\chi(n+N)(n+N-1)c_{2}(t))\mid
n+N,b>),
\end{eqnarray}
comparing the both sides of Eq. (5), we have
\begin{eqnarray}
i\frac{\partial}{\partial t}c_1(t)=\omega
nc_{1}(t)+\frac{1}{2}\omega_0c_{1}(t)+gc_{2}(t)\sqrt{(n+1)(n+2)\cdot\cdot\cdot(n+N)}+\chi
n(n-1)c_{1}(t),
\end{eqnarray}

\begin{eqnarray}
i\frac{\partial}{\partial t}c_2(t)=\omega
(n+N)c_{2}(t)-\frac{1}{2}\omega_0c_{2}(t)+gc_{1}(t)\sqrt{(n+2)(n+1)\cdot\cdot\cdot(n+N)}+\chi(n+N)(n+N-1)c_{2}(t),
\end{eqnarray}

using Laplace transforms to the both sides of Eqs. (6) and (7), we
get
\begin{eqnarray}
ipL_{1}(p)-ic_{1}(0)=\omega
nL_{1}(P)+\frac{1}{2}\omega_{0}L_{1}(p)+g\sqrt{(n+1)(n+2)\cdot\cdot\cdot(n+N)}L_{2}(p)+\chi
n(n-1)L_{1}(p),
\end{eqnarray}
\begin{eqnarray}
ipL_{2}(p)-ic_{2}(0)&&=\omega
(n+N)L_{2}(P)-\frac{1}{2}\omega_{0}L_{2}(p)+g\sqrt{(n+1)(n+2)\cdot\cdot\cdot(n+N)}L_{1}(p)
\nonumber\\&&+\chi(n+N)(n+N-1)L_{2}(p),
\end{eqnarray}

We obtain
\begin{eqnarray}
&&L_{1}(p)=\frac{c_{1}}{p+i\omega n+\frac{i}{2}\omega_{0}+i\chi
n(n-1)}-i\frac{g\sqrt{(n+1)(n+2)\cdot\cdot\cdot(n+N)}}{p+i\omega
n+\frac{i}{2}\omega_{0}+i\chi
n(n-1)}(\frac{D}{p+A}+\frac{E}{p+B}),
\end{eqnarray}
\begin{eqnarray}
L_{2}(p)=\frac{D}{p+A}+\frac{E}{p+B},
\end{eqnarray}

where $L_1(p)={\cal{L}}[c_{1}(t)]$ and
$L_2(p)={\cal{L}}[c_{2}(t)]$ are Laplace transform of functions
$c_{1}(t)$, $c_{2}(t)$, and

\begin{eqnarray}
A=\frac{i\omega(2n+N)+i\chi[2n^{2}+2(N-1)n+N^{2}-N]+i\sqrt{\omega_2^2+\omega_1^2}}{2},
\end{eqnarray}
\begin{eqnarray}
B=\frac{i\omega(2n+N)+i\chi[2n^{2}+2(N-1)n+N^{2}-N]-i\sqrt{\omega_2^2+\omega_1^2}}{2},
\end{eqnarray}
\begin{eqnarray}
D=\frac{(\sqrt{\omega_1^2+\omega_2^2}-\omega_2)c_2+\omega_1
c_1}{2\sqrt{\omega_1^2+\omega_2^2}},
E=\frac{(\sqrt{\omega_1^2+\omega_2^2}-\omega_2)c_2-\omega_1
c_1}{2\sqrt{\omega_1^2+\omega_2^2}},
\end{eqnarray}
Using Laplace retransforms to Eqs. (10) and (11), we have

\begin{eqnarray}
c_{1}(t)=e^{-\frac{i}{2}\omega(2n+N)t}e^{-\frac{i}{2}\chi[2n^{2}+2(N-1)n+N^{2}-N]t}[c_{1}\cos(\frac{\sqrt{\omega_1^2+\omega_2^2}}{2}t)
-i\frac{c_1\omega_2+c_2\omega_1}{\sqrt{\omega_1^2+\omega_2^2}}\sin(\frac{\sqrt{\omega_1^2+\omega_2^2}}{2}t)]
\end{eqnarray}

\begin{eqnarray}
c_{2}(t)=e^{-\frac{i}{2}\omega(2n+N)t}e^{-\frac{i}{2}\chi[2n^{2}+2(N-1)n+N^{2}-N]t}[c_{2}\cos(\frac{\sqrt{\omega_1^2+\omega_2^2}}{2}t)
-i\frac{c_1\omega_1-c_2\omega_2}{\sqrt{\omega_1^2+\omega_2^2}}\sin(\frac{\sqrt{\omega_1^2+\omega_2^2}}{2}t)]
\end{eqnarray}
where $\delta=\omega_0-N\omega$,
$\omega_1=2g\sqrt{(n+1)(n+2)\cdot\cdot\cdot(n+N)}$,
$\omega_2=\delta-N\chi(2n+N-1)$.

with the state (2), we can obtain the density operator of
atom-photon system
\begin{eqnarray}
\hat{\rho}(t)&&=|\psi(t)><\psi(t)|\nonumber\\&&=|c_1(t)|^2|n,a><a,n|+c_1(t)c_2^{*}(t)|n,a><b,n+2|
\nonumber\\&&+c_2(t)c_1^{*}(t)|n+2,b><a,n|+|c_2(t)|^2|n+2,b><b,n+2|,
\end{eqnarray}
the reduce density operator of atom $A$ is
\begin{eqnarray}
\hat{\rho}_A(t)&&=tr_{(a)}\hat{\rho}(t)
\nonumber\\&&=<n|\hat{\rho}(t)|n>+<n+2|\hat{\rho}(t)|n+2>
\nonumber\\&&=|c_1(t)|^2|a><a|+|c_2(t)|^2|b><b|,
\end{eqnarray}
the matrix form of $\hat{\rho}_A(t)$ at basis vectors $|a>$ and
$|b>$
\begin{eqnarray}
\hat{\rho}_A(t)=\left(
\begin{array}{cc}
<a|\hat{\rho}_A(t)|a>& <a|\hat{\rho}_A(t)|b> \\
<b|\hat{\rho}_A(t)|a> & <b|\hat{\rho}_A(t)|b>%
\end{array}%
\right)=\left(
\begin{array}{cc}
|c_1(t)|^2& 0 \\
0 & |c_2(t)|^2%
\end{array}%
\right),
\end{eqnarray}
the quantum system entanglement degrees is
\begin{eqnarray}
E(t)&&=-tr (\hat{\rho}_A(t)\log_2\hat{\rho}_A(t))
\nonumber\\&&=-tr(\left(
\begin{array}{cc}
|c_1(t)|^2& 0 \\
0 & |c_2(t)|^2%
\end{array}%
\right)\cdot\left(
\begin{array}{cc}
|\log_2c_1(t)|^2& 0 \\
0 & \log_2|c_2(t)|^2%
\end{array}%
\right)),
\nonumber\\&&=-(|c_1(t)|^2\log_2c_1(t)|^2+|c_2(t)|^2\log_2|c_2(t)|^2),
\end{eqnarray}

 \vskip 8pt {\bf 3. Numerical result} \vskip 8pt

In this section, we shall calculate the quantum entanglement
degrees with Eqs. (15), (16) and (20). The entanglement degree $E$
is in the range of $0\sim 1$. From FIGs. 1 to 5, we calculate the
entanglement degree with linear Jaynes-Cummings model, i.e.,
$\chi=0$. In FIGs. 1 and 2, the initial state superposition
coefficient $C_{1}=0$, the entanglement degree $E=0$, i.e., the
atom and light field is not in entangled state at initial state,
the initial photon number $n=0$. With the time evolution, the atom
and light field should be in the entangled state $E\neq 0$. In
FIG. 1 (a)-(f), the quantum discord $\delta^{2}=0$, the transition
photon numbers $N$ are $1, 2, 3, 4, 5$ and $6$, respectively. In
FIG.1 (a)-(f), the entanglement degree $0 \leq E \leq 1$. When the
numbers of photons $N$ increases, the entanglement degrees
oscillation get faster. When $N=1, 2, 3$, the evolution curves of
entanglement degrees change slowly with time $t$, and stay the
time of entanglement degrees $E\approx 1$ more longer. When $N=4,
5, 6$, the entanglement degree oscillate quickly, and they
oscillate more quickly with $N$ increase. In FIG. 2, the quantum
discord $\delta^{2}=4g^{2}$, and other parameters are the same as
FIG. 1. Comparing FIG. 2 with FIG. 1, When the quantum discord
$\delta$ increase, the entanglement degrees oscillation get
slowly. In FIG. 3, the initial state superposition coefficient
$C_{1}=0.96$, and other parameters are the same as FIG. 1. The
atom and light field is in entangled state at initial state. With
the time evolution, the atom and light field should be in the
entangled state. When $N$ increase, the entangled degrees increase
for $N=1, 2, 3, 4, 5, 6$. In FIGs. 4 and 5, the initial state
superposition coefficient $C_{1}=0.76$, the atom and light field
is in maximum entangled state at initial state. In FIG. 4, the
quantum discord $\delta^{2}=0$. With the time evolution, the atom
and light field should be in the maximum entangled state for $N=1,
2, 3, 4, 5, 6$. In FIG. 5, the quantum discord
$\delta^{2}=8g^{2}$. With the quantum discord $\delta$ increase,
when $N=1, 2, 3$, the entanglement degree oscillate quickly, when
$N=4, 5, 6$, they keep the maximum entangled state.

From FIGs. 6 to 9, we calculate the entanglement degree with
nonlinear Jaynes-Cummings model, i.e., $\chi\neq 0$. The initial
state superposition coefficient $C_{1}=0$, the initial photon
number $n=0$, the quantum discord $\delta^{2}=4g^{2}$, and they
corresponding to nonlinear coefficient are
$\alpha=\frac{\chi}{g}=0.2$, $0.8$, $-0.2$ and $-0.8$. Comparing
FIGs. 6 with 7, we find when the nonlinear coefficient increase (
$\alpha >0$), the entanglement degrees oscillation get quickly.
Comparing FIGs. 8 with 9, we find when the nonlinear coefficient
decrease £¨$\alpha <0$£©, the entanglement degrees oscillation get
quickly. Comparing FIG. 6 with 2, we find when the nonlinear
coefficient $\alpha >0$, the entanglement degrees oscillation get
quickly, to the disadvantage of the atom and light field
entanglement. Comparing FIG. 8 with 2, we find when the nonlinear
coefficient $\alpha <0$, the entanglement degrees oscillation get
slow. It is benefit to the atom and light field entanglement
obviously.
 \vskip 8pt {\bf 4. Conclusion} \vskip 8pt

In this paper, we have studied the atom and light field quantum
entanglement of multiphoton transition, and researched the effect
of initial state superposition coefficient $C_{1}$, the initial
photon number $n$, the transition photon number $N$ ($N=1, 2, 3,
4, 5, 6$), the quantum discord $\delta$ and nonlinear coupling
constant $\chi$ on the quantum entanglement degrees. We have given
the quantum entanglement degrees curves with time evolution, and
obtained some results. When the transition photon number $N$
increases, the entanglement degrees oscillation get faster. When
the initial state superposition coefficient $C_{1}=0$,  with the
quantum discord $\delta$ increase, the entanglement degrees
oscillation get slowly. When the initial state superposition
coefficient $C_{1}=0.76$, the the quantum discord increases, the
entanglement degrees oscillation get quickly. when the nonlinear
coefficient $\alpha >0$, the entanglement degrees oscillation get
quickly. When the nonlinear coefficient $\alpha <0$, the
entanglement degrees oscillation get slow. It is benefit to the
atom and light field entanglement obviously. These results have
important significance in the quantum communication and quantum
information.

\vskip 12pt {\bf 5.  Acknowledgment} \vskip 12pt

This work is supported by Scientific and Technological Development
Foundation of Jilin Province, Grant Number: 20130101031JC.

\newpage
\begin{figure}[tbp]
\includegraphics[width=16cm, height=8cm]{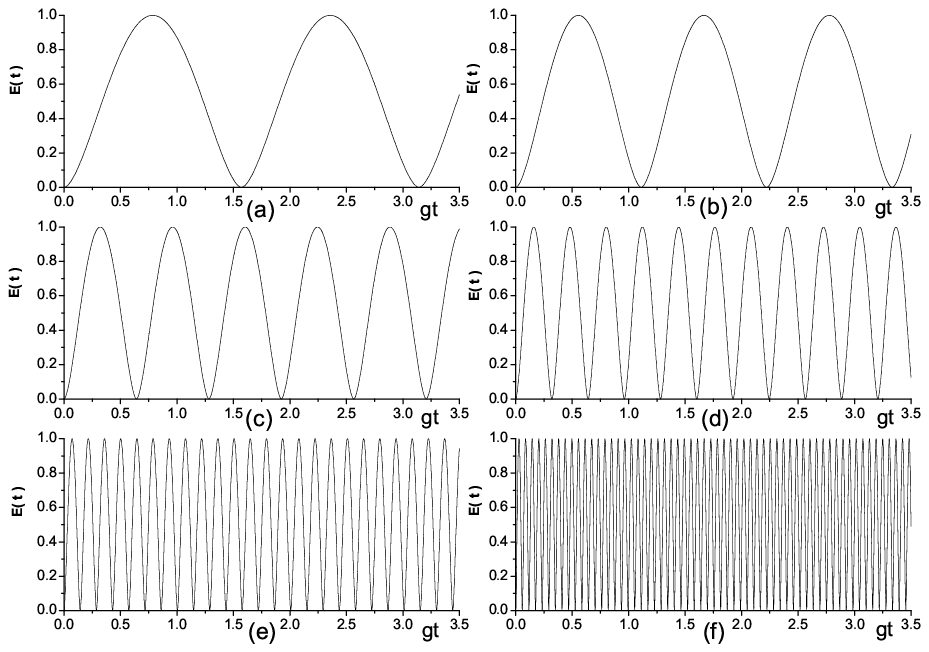}
\caption{The curves of the atom and light field entanglement
degrees with time evolution, the initial state superposition
coefficient $C_{1}=0$, the initial photon number $n=0$, the
quantum discord $\delta^{2}=0$, the nonlinear coefficient
$\alpha=\frac{\chi}{g}=0$.}
\end{figure}
\begin{figure}[tbp]
\includegraphics[width=16cm, height=8cm]{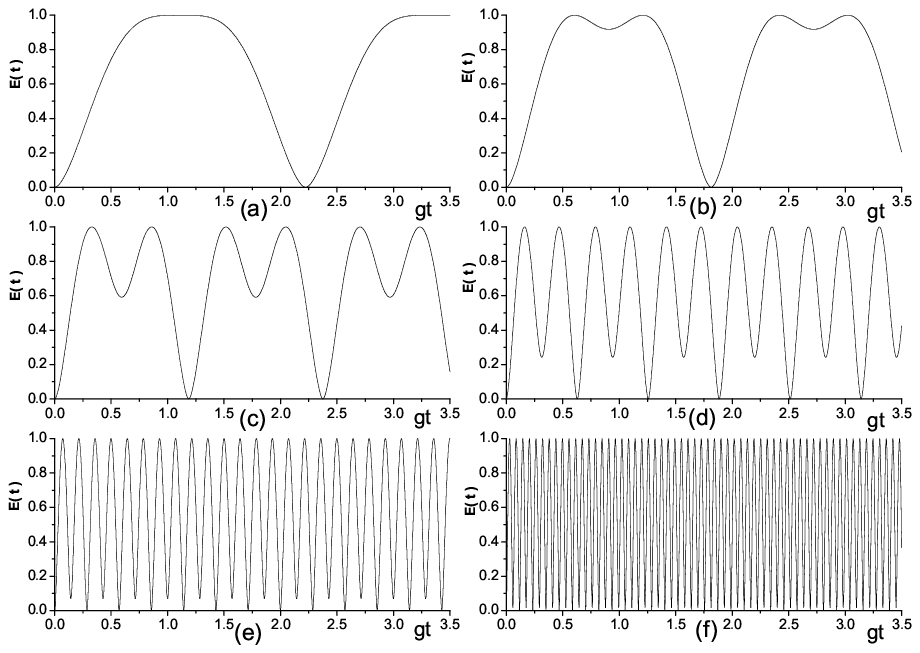}
\caption{The curves of the atom and light field entanglement
degrees with time evolution, the initial state superposition
coefficient $C_{1}=0$, the initial photon number $n=0$, the
quantum discord $\delta^{2}=4g^{2}$, the nonlinear coefficient
$\alpha=\frac{\chi}{g}=0$.}
\end{figure}
\begin{figure}[tbp]
\includegraphics[width=16cm, height=8cm]{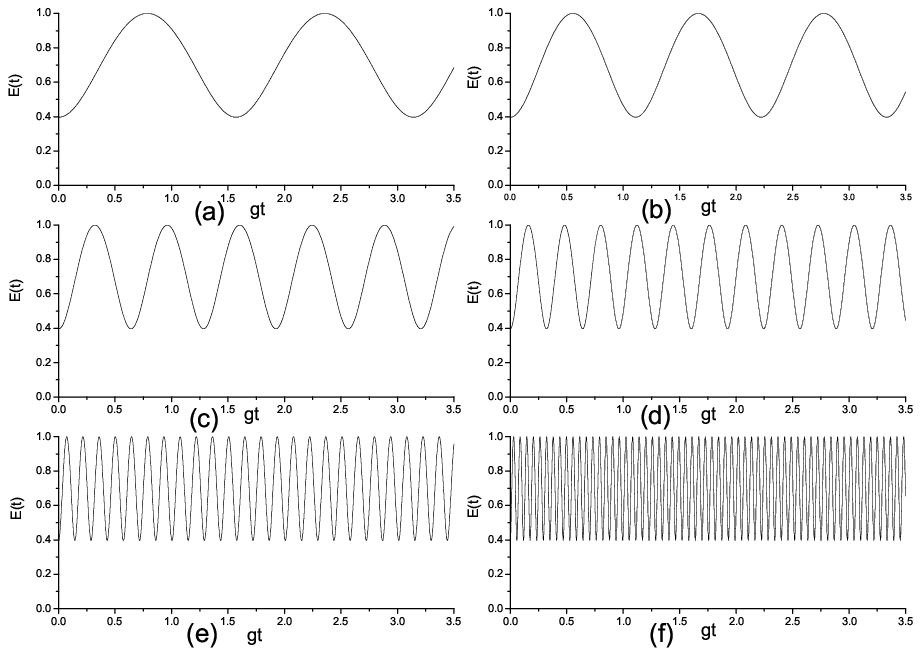}
\caption{The curves of the atom and light field entanglement
degrees with time evolution, the initial state superposition
coefficient $C_{1}=0.96$, the initial photon number $n=0$, the
quantum discord $\delta^{2}=0$, the nonlinear coefficient
$\alpha=\frac{\chi}{g}=0$.}
\end{figure}
\begin{figure}[tbp]
\includegraphics[width=16cm, height=8cm]{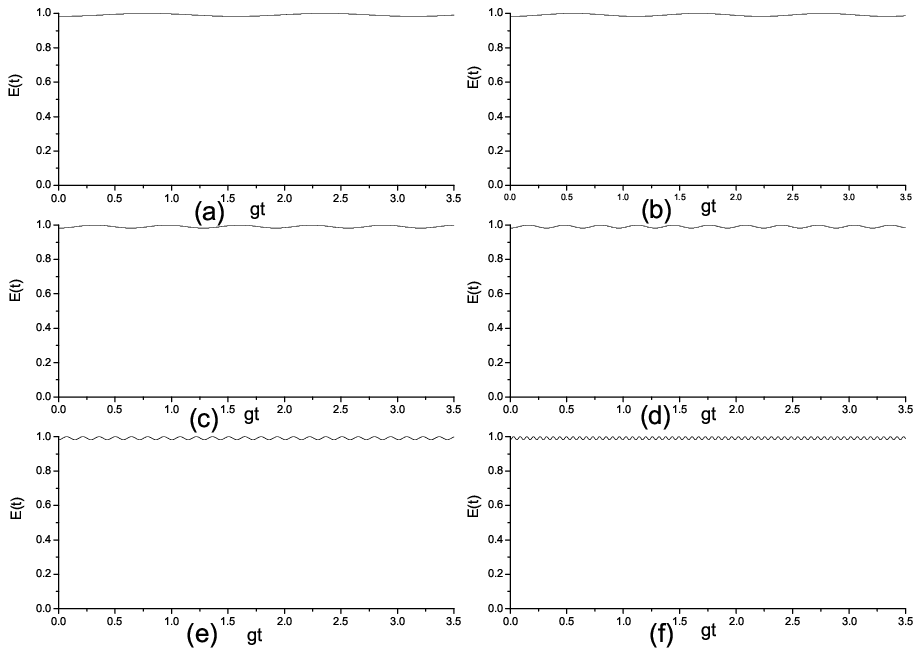}
\caption{The curves of the atom and light field entanglement
degrees with time evolution, the initial state superposition
coefficient $C_{1}=0.76$, the initial photon number $n=0$, the
quantum discord $\delta^{2}=0$, the nonlinear coefficient
$\alpha=\frac{\chi}{g}=0$.}
\end{figure}
\begin{figure}[tbp]
\includegraphics[width=16cm, height=8cm]{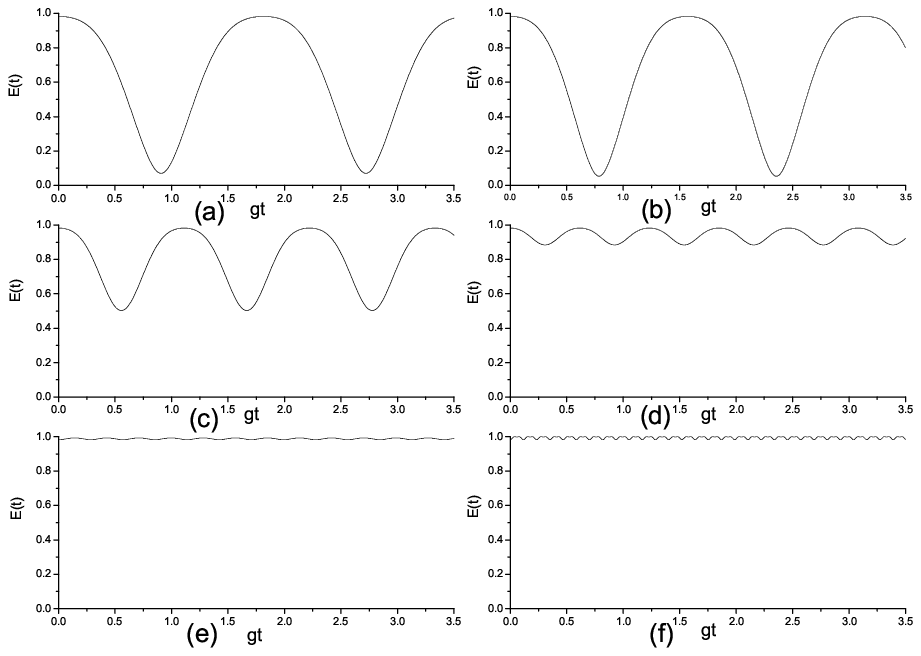}
\caption{The curves of the atom and light field entanglement
degrees with time evolution, the initial state superposition
coefficient $C_{1}=0.76$, the initial photon number $n=0$, the
quantum discord $\delta^{2}=8g^{2}$, the nonlinear coefficient
$\alpha=\frac{\chi}{g}=0$.}
\end{figure}
\begin{figure}[tbp]
\includegraphics[width=16cm, height=8cm]{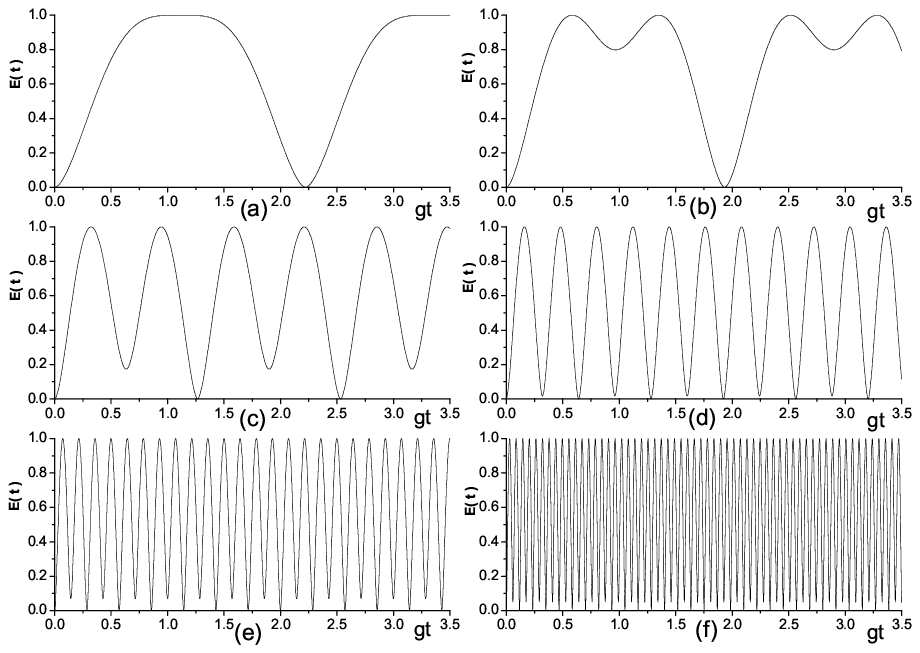}
\caption{The curves of the atom and light field entanglement
degrees with time evolution, the initial state superposition
coefficient $C_{1}=0$, the initial photon number $n=0$, the
quantum discord $\delta^{2}=4g^{2}$, the nonlinear coefficient
$\alpha=\frac{\chi}{g}=0.2$.}
\end{figure}
\begin{figure}[tbp]
\includegraphics[width=16cm, height=8cm]{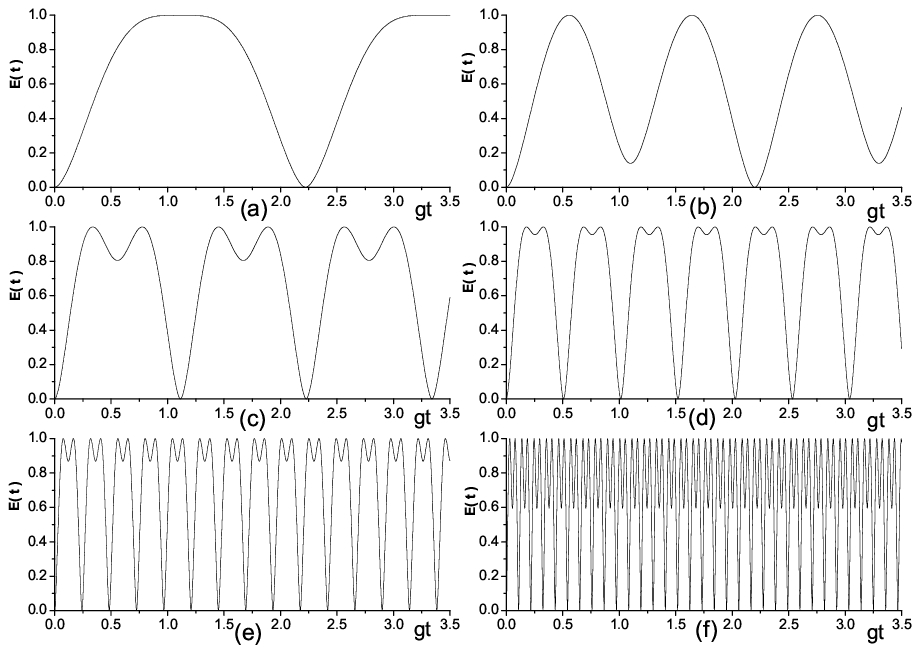}
\caption{The curves of the atom and light field entanglement
degrees with time evolution, the initial state superposition
coefficient $C_{1}=0$, the initial photon number $n=0$, the
quantum discord $\delta^{2}=4g^{2}$, the nonlinear coefficient
$\alpha=\frac{\chi}{g}=0.8$.}
\end{figure}
\begin{figure}[tbp]
\includegraphics[width=16cm, height=8cm]{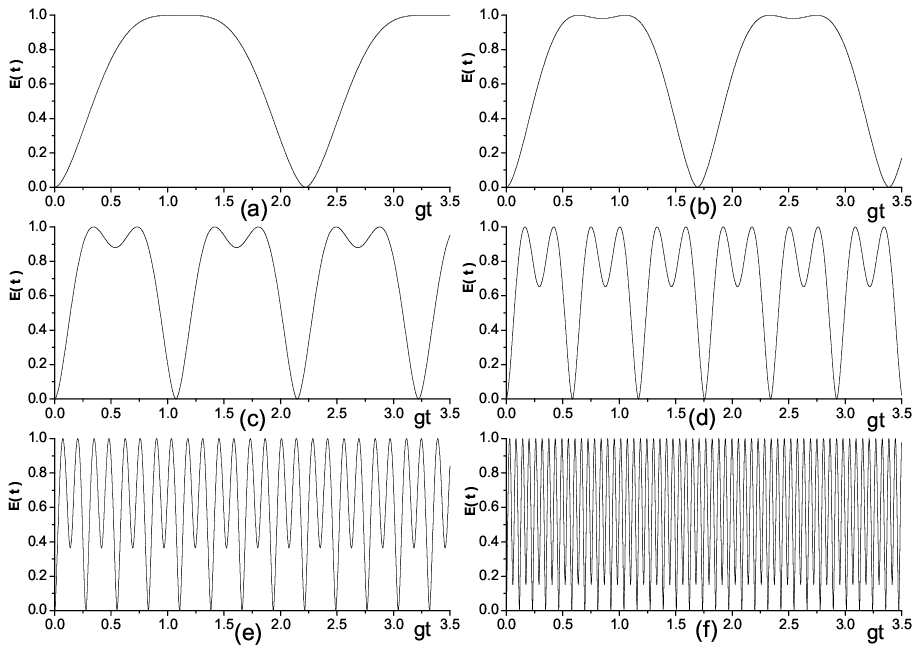}
\caption{The curves of the atom and light field entanglement
degrees with time evolution, the initial state superposition
coefficient $C_{1}=0$, the initial photon number $n=0$, the
quantum discord $\delta^{2}=4g^{2}$, the nonlinear coefficient
$\alpha=\frac{\chi}{g}=-0.2$.}
\end{figure}
\begin{figure}[tbp]
\includegraphics[width=16cm, height=8cm]{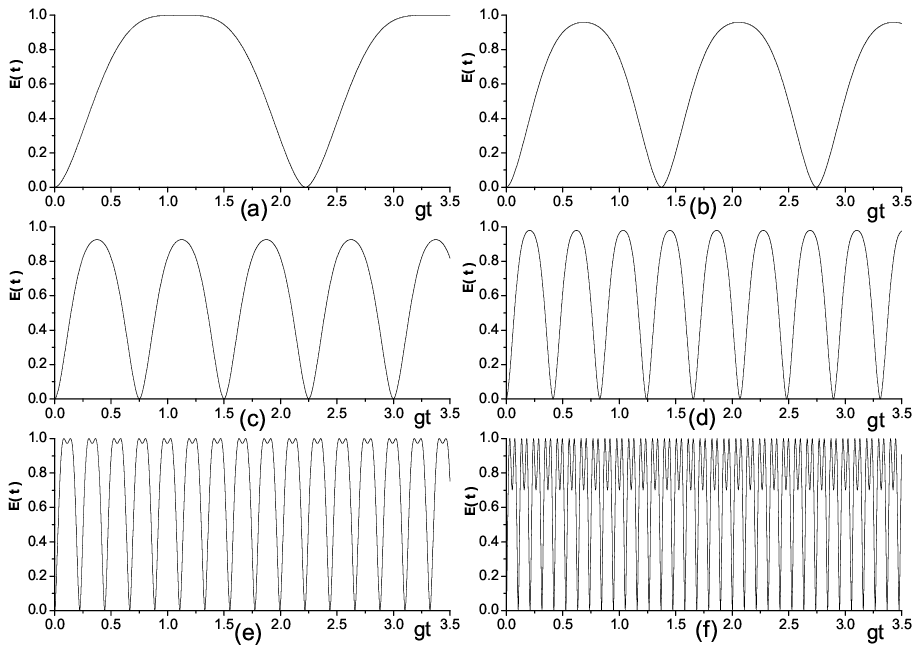}
\caption{The curves of the atom and light field entanglement
degrees with time evolution, the initial state superposition
coefficient $C_{1}=0$, the initial photon number $n=0$, the
quantum discord $\delta^{2}=4g^{2}$, the nonlinear coefficient
$\alpha=\frac{\chi}{g}=-0.8$.}
\end{figure}

\end{document}